\documentclass[pra,showpacs,amsmath,twocolumn]{revtex4}
\usepackage{revsymb}
\usepackage{graphicx}
\usepackage{bm}

\begin{document}
\newcommand{\ket}[1]{|{#1}\rangle}
\newcommand{\bra}[1]{\langle{#1}|}
\newcommand{\bras}[2]{{}_{#2}\hspace*{-0.2truemm}\langle{#1}|}
\newcommand{\ketbras}[3]{\ket{#1}_{#3}\hspace*{-0.2truemm}\bra{#2}}
\newcommand{\brackets}[3]{{}_{#3}\hspace*{-0.2mm}\langle{#1}|{#2}\rangle_{#3}}
\newcommand{\Tr}{\mathop{\text{Tr}}\nolimits}

\preprint{WU--HEP--04--01}
\title{Preparation and entanglement purification of qubits
through Zeno-like measurements}
\author{Hiromichi Nakazato}
\email{hiromici@waseda.jp}
\author{Makoto Unoki}
\author{Kazuya Yuasa}
\email{yuasa@hep.phys.waseda.ac.jp}
\affiliation{Department of Physics, Waseda University, Tokyo
169-8555, Japan}
\date[]{April 16, 2004}
\begin{abstract}
A novel method of purification, \textit{purification through
Zeno-like measurements} [H. Nakazato, T. Takazawa, and K. Yuasa,
Phys.\ Rev.\ Lett.\ \textbf{90}, 060401 (2003)], is discussed
extensively and applied to a few simple qubit systems.
It is explicitly demonstrated how it works and how it is
optimized.
As possible applications, schemes for \textit{initialization of
multiple qubits} and \textit{entanglement purification} are
presented, and their efficiency is investigated in detail.
Simplicity and flexibility of the idea allow us to apply it to
various kinds of settings in quantum information and computation,
and would provide us with useful and practical methods of state
preparation.
\end{abstract}
\maketitle

\section{Introduction}
It is usually not seriously discussed in normal textbooks on
quantum mechanics about \textit{how to prepare an initial state}.
It is, however, becoming an important subject not only from a
view point of foundation of quantum mechanics, but also from a
practical point of view, since we are rushing towards
experimental realizations of the ideas for quantum information
and computation \cite{ref:QuantInfoCompChuang,%
ref:QuantInfoCompZeilinger}.
Without establishing particular initial states assumed in several
algorithms, we cannot start any processes of the attractive
ideas.
State preparation is one of the key elements to quantum
information processing \cite{ref:QuantInfoCompChuang,%
ref:QuantInfoCompZeilinger}, and there are several theoretical
proposals \cite{ref:StatePreparationTh,%
ref:EntanglementPreparationTh,ref:EntanglementSeparate} and
experimental attempts \cite{ref:Haroche1997,ref:NIST,ref:Blatt,%
ref:NEC,ref:YamamotoNumberState2003}.

In the ideas for quantum information and computation, quantum
states with high coherence, especially \textit{entangled states},
play significant and essential roles.
But such ``clean'' states required for quantum information
technologies are not easily found in nature, since many of them
are fragile against environmental perturbations and suffer from
\textit{decoherence}.
Therefore, there would often be a demand for preparing a desired
\textit{pure state} out of an arbitrary \textit{mixed state}.
Several schemes have been proposed for it, which are called
``purification,'' ``distillation,'' ``concentration,''
``extraction,'' etc.~\cite{ref:QuantInfoCompZeilinger,%
ref:PurificationBennett,ref:PurificationExperiments}.

One of the simplest and easiest ways of state preparation is to
resort to a projective measurement:  a quantum system shall be in
a pure state $\ket{\phi}$ after it is measured and confirmed to
be in the state $\ket{\phi}$.
Such a strategy is not possible, however, in cases where the
desired state $\ket{\phi}$ cannot be directly measured or where
the relevant system is not available after the confirmation.
This is often the case for entangled states, which are the key
resources to quantum information and computation.
This is why more elaborate purification protocols are required
and several schemes of \textit{entanglement purification/
preparation} have been proposed
\cite{ref:QuantInfoCompZeilinger,ref:PurificationBennett,%
ref:PurificationExperiments}.

Recently, a novel mechanism to purify quantum states has been
found and reported: \textit{purification through Zeno-like
measurements} \cite{ref:qpf}.
A pure state is extracted in a quantum system through a series of
repeated measurements (Zeno-like measurements) on another quantum
system in interaction with the former.
Since the relevant system to be purified is not directly measured
in this scheme, it would be suitable for such situations
mentioned above.
In this article, we discuss this scheme in detail and explore, on a heuristic basis, its potential as a useful and effective method of purification of qubits.
The examples considered here are quite simple but still possess
potential and practical applicability.

This article is organized as follows.
First, the basic framework of the purification is described in a
general setting, and the conditions for the purification and its
optimization are summarized in Sec.~\ref{sec:Framework}, where
some details which are not discussed in the first report
\cite{ref:qpf} are included.
It is then demonstrated in Sec.~\ref{sec:Single} how it works and
how it can be made optimal in a simplest example, i.e.,
\textit{single-qubit purification}, and a generalization to a
multi-qubit case is considered in Sec.~\ref{sec:Initialization},
which would afford us a useful method of \textit{initialization
of multiple qubits}.
One of the interesting applications of the present scheme is
\textit{entanglement purification}, which is discussed in
Sec.~\ref{sec:EntanglementPurification} and shown to be actually
possible.
Concluding remarks are given in Sec.~\ref{sec:Summary} with some
comments on possible extensions and future subjects.
Appendices A--E are supplied in order to demonstrate detailed
calculations and proofs, that are not described in the text.

\section{Framework}
\label{sec:Framework}
\begin{figure}
\includegraphics[width=0.3\textwidth]{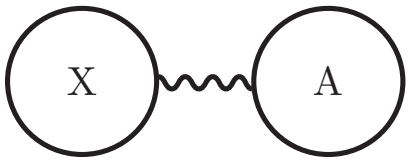}
\caption{We repeat measurements on X and purify A\@.}
\label{fig:CoupledSystem}
\end{figure}

Let us recapitulate the framework of the purification reported in
\cite{ref:qpf}.
We consider two quantum systems X and A interacting with each
other (Fig.~\ref{fig:CoupledSystem}).
The total system X+A is initially in a \textit{mixed} state
$\varrho_\text{tot}$, from which we try to extract a pure state
in A by controlling X\@.
We first perform a measurement on X (the zeroth measurement)
\textit{to confirm that it is in a state $\ket{\phi}_\text{X}$}.
If it is found in the state $\ket{\phi}_\text{X}$, the state of
the total system is projected by the projection operator
\begin{equation}
\mathcal{O}
=\ketbras{\phi}{\phi}{\text{X}}\otimes\openone_\text{A}
\end{equation}
to yield
\begin{equation}
\varrho_\text{tot}
\to\tilde{\varrho}_\text{tot}
=\frac{\mathcal{O}\varrho_\text{tot}\mathcal{O}}%
{\Tr(\mathcal{O}\varrho_\text{tot}\mathcal{O})}
=\ketbras{\phi}{\phi}{\text{X}}\otimes\varrho_\text{A},
\end{equation}
where $\varrho_\text{A}\equiv\bras{\phi}{\text{X}}%
\varrho_\text{tot}\ket{\phi}_\text{X}/P_0$ is the state of A
after this zeroth confirmation and $P_0\equiv\Tr(\mathcal{O}%
\varrho_\text{tot}\mathcal{O})$ is the probability for this to
happen.
We then let the total system start to evolve under a total
Hamiltonian $H_\text{tot}$ and repeat the same measurement on X
at regular time intervals $\tau$.
After $N$ repetitions of successful confirmations, i.e., after X
is confirmed to be in the state $\ket{\phi}_\text{X}$
\textit{successively} $N$ times, the state of the total system,
$\varrho_\text{tot}^{(\tau)}(N)$, is cast into the following
form:
\begin{subequations}
\label{eqn:State}
\begin{align}
\varrho_\text{tot}^{(\tau)}(N)
&=(\mathcal{O}e^{-iH_\text{tot}\tau})^N
\tilde{\varrho}_\text{tot}
(e^{iH_\text{tot}\tau}\mathcal{O})^N/\tilde{P}^{(\tau)}(N)
\nonumber\displaybreak[0]\\
&=\ketbras{\phi}{\phi}{\text{X}}\otimes
\varrho_\text{A}^{(\tau)}(N),
\label{eqn:StateTotal}\displaybreak[0]\\
\varrho_\text{A}^{(\tau)}(N)
&=\bm{(}V_\phi(\tau)\bm{)}^N\varrho_\text{A}
\bm{(}V_\phi^\dag(\tau)\bm{)}^N/\tilde{P}^{(\tau)}(N),
\label{eqn:StateA}
\end{align}
\end{subequations}
where $V_\phi(\tau)$, defined by
\begin{equation}
V_\phi(\tau)
\equiv\bras{\phi}{\text{X}}e^{-iH_\text{tot}\tau}
\ket{\phi}_\text{X},
\label{eqn:V}
\end{equation}
is a projected time-evolution operator acting on the Hilbert
space of A, and $\tilde{P}^{(\tau)}(N)$ is the normalization
factor,
\begin{align}
\tilde{P}^{(\tau)}(N)
&=\Tr[(\mathcal{O}e^{-iH_\text{tot}\tau})^N
\tilde{\varrho}_\text{tot}(e^{iH_\text{tot}\tau}\mathcal{O})^N]
\nonumber\displaybreak[0]\\
&=\Tr_\text{A}[\bm{(}V_\phi(\tau)\bm{)}^N\varrho_\text{A}
\bm{(}V_\phi^\dag(\tau)\bm{)}^N].
\label{eqn:Yield}
\end{align}
Note that we retain only those events where X is found in the
state $\ket{\phi}_\text{X}$ at \textit{every} measurement
(including the zeroth one); other events, resulting in failure to
purify A, are discarded.
The normalization factor $\tilde{P}^{(\tau)}(N)$ multiplied by
$P_0$, i.e., $P^{(\tau)}(N)\equiv\tilde{P}^{(\tau)}(N)P_0$, is
nothing but the probability for the \textit{successful events}
and is the probability of obtaining the state given in
(\ref{eqn:State}).

For definiteness, let us restrict ourselves on finite-dimensional systems throughout this article and consider the spectral decomposition of the operator $V_\phi(\tau)$.
Since the operator $V_\phi(\tau)$ is not a Hermitian operator, we should set up both right and left eigenvalue equations
\begin{subequations}
\begin{align}
V_\phi(\tau)\ket{u_n}_\text{A}&=\lambda_n\ket{u_n}_\text{A},\\
\bras{v_n}{\text{A}}V_\phi(\tau)&=\lambda_n\,\bras{v_n}{\text{A}}.
\end{align}
\end{subequations}
The eigenvalues $\lambda_n$ are complex in general and bounded as
\begin{equation}
0\le|\lambda_n|\le1
\label{eqn:Bound}
\end{equation}
(see Appendix \ref{app:Bound}).
Here we assume for simplicity that the spectrum of the operator $V_\phi(\tau)$ is not degenerate.
In such a case, the eigenvectors are orthogonal to each other in the sense
\begin{subequations}
\begin{equation}
\brackets{v_m}{u_n}{\text{A}}=\delta_{mn}
\end{equation}
and form a complete set in the Hilbert space of system A,
\begin{equation}
\sum_n\ketbras{u_n}{v_n}{\text{A}}=\openone_\text{A},
\end{equation}
\end{subequations}
which readily leads to the spectral decomposition of the operator $V_\phi(\tau)$,
\begin{equation}
V_\phi(\tau)=\sum_n\lambda_n\ketbras{u_n}{v_n}{\text{A}}.
\label{eqn:SpectralDecomp}
\end{equation}
(In the following, we also normalize the right eigenvectors as $\brackets{u_n}{u_n}{\text{A}}=1$.)

Even in a general situation where the spectrum of the operator $V_\phi(\tau)$ is degenerate, the diagonalization (\ref{eqn:SpectralDecomp}) is possible when and only when all the right eigenvectors $\ket{u_n}_\text{A}$ are linearly independent of each other and form a complete basis \cite{ref:Kato}.
Otherwise, the spectral decomposition is not like (\ref{eqn:SpectralDecomp}), but in the ``Jordan canonical form'' \cite{ref:Kato}.
The diagonalizability of the operator $V_\phi(\tau)$ is, however, not an essential assumption as clarified in Appendix \ref{app:JordanDecomposition}\@.

It is now easy to observe the asymptotic behavior of the state of
A, $\varrho_\text{A}^{(\tau)}(N)$ in (\ref{eqn:StateA}).
Since the eigenvalues $\lambda_n$ are bounded like (\ref{eqn:Bound}), each term in the expansion
\begin{equation}
\bm{(}V_\phi(\tau)\bm{)}^N
=\sum_n\lambda_n^N\ketbras{u_n}{v_n}{\text{A}}
\end{equation}
decays out and a single term dominates asymptotically as the
number of measurements, $N$, increases,
\begin{equation}
\bm{(}V_\phi(\tau)\bm{)}^N
\to\lambda_0^N\ketbras{u_0}{v_0}{\text{A}}
\quad\mbox{as $N$ increases},
\label{eqn:Mechanism}
\end{equation}
\textit{provided}
\begin{multline}
\text{\textit{the largest (in magnitude) eigenvalue $\lambda_0$
is}}\\[-3pt]
\text{\textit{unique, discrete and non\-degenerate}}.
\label{eqn:Condition}
\end{multline}
[The word ``unique'' means that there is only one eigenvalue that has the maximum modulus and ``nondegenerate'' means that there is only one right eigenvector (and a corresponding left eigenvector) belonging to that maximal (in magnitude) eigenvalue.]
Thus, the state of A in (\ref{eqn:StateA}) approaches a pure
state $\ket{u_0}_\text{A}$,
\begin{equation}
\varrho_\text{A}^{(\tau)}(N)\to\ketbras{u_0}{u_0}{\text{A}}
\quad\mbox{as}\quad N\to\infty.
\label{eqn:Purification}
\end{equation}
This is the purification scheme proposed recently \cite{ref:qpf}:
extraction of a pure state $\ket{u_0}_\text{A}$ through a series
of repeated measurements on X\@.
Since we repeat measurements (on X) as in the case of the quantum
Zeno effect \cite{ref:QZE}, we call such measurements ``Zeno-like
measurements'' \cite{note:QZE}.
The final pure state $\ket{u_0}_\text{A}$ is the eigenstate of
the projected time-evolution operator $V_\phi(\tau)$ belonging to
the largest (in magnitude) eigenvalue $\lambda_0$ and depends on
the parameters $\tau$, $\ket{\phi}_\text{X}$, and those in the
Hamiltonian $H_\text{tot}$.
It is, however, independent of the initial state
$\varrho_\text{tot}$.
The pure state $\ket{u_0}_\text{A}  $ is extracted from an
\textit{arbitrary} mixed state $\varrho_\text{tot}$ through the
Zeno-like measurements.
By tuning such parameters mentioned above, we have a possibility
of extracting a desired pure state $\ket{u_0}_\text{A}$.

The above observation shows that the assumption of the diagonalizability in (\ref{eqn:SpectralDecomp}) is not essential but condition (\ref{eqn:Condition}), i.e., the existence of the \textit{unique, discrete and nondegenerate largest (in magnitude) eigenvalue $\lambda_0$}, is crucial to the purification.
For our purification mechanism to work, it is crucial that a single state is extracted and this is accomplished when these qualifications, i.e., the uniqueness of the largest eigenvalue and the nondegeneracy of the eigenvector, are both met.
The diagonalizability of $V_\phi(\tau)$ is not relevant to these conditions and is not essential to the purification.
This point is clarified in Appendix \ref{app:JordanDecomposition}\@.

Furthermore, note the asymptotic behavior of the success
probability $P^{(\tau)}(N)$: it decays asymptotically as
\begin{align}
P^{(\tau)}(N)
\to{}&|\lambda_0|^{2N}P_0\,\bras{v_0}{\text{A}}\varrho_\text{A}
\ket{v_0}_\text{A}\nonumber\\
={}&|\lambda_0|^{2N}\bras{\phi v_0}{\text{XA}}\varrho_\text{tot}
\ket{\phi v_0}_\text{XA}\quad\mbox{as $N$ increases},
\label{eqn:Decay}
\end{align}
where $\ket{\phi u_0}_\text{XA}$ stands for $\ket{\phi}_\text{X}%
\otimes\ket{u_0}_\text{A}$ and $\bras{\phi u_0}{\text{XA}}%
=\bras{\phi}{\text{X}}\otimes\bras{u_0}{\text{A}}$.
The decay is governed by the eigenvalue $\lambda_0$, and
therefore, an efficient purification is possible if $\lambda_0$
satisfies the condition
\begin{equation}
|\lambda_0|=1,
\label{eqn:OptimizationI}
\end{equation}
which suppresses the decay in (\ref{eqn:Decay}) to give the final
(nonvanishing) success probability
\begin{equation}
P^{(\tau)}(N)
\to\bras{\phi v_0}{\text{XA}}\varrho_\text{tot}
\ket{\phi v_0}_\text{XA}.
\end{equation}
It is worth stressing that the condition
(\ref{eqn:OptimizationI}) allows us to repeat the measurement as
many times as we wish without running the risk of losing the
success probability $P^{(\tau)}(N)$.
In other words, high fidelity to the target state and
nonvanishing success probability do not contradict each other in
this scheme, but rather they can be achieved simultaneously.
At the same time, if the other eigenvalues are much smaller than
$\lambda_0$ in magnitude,
\begin{equation}
|\lambda_n/\lambda_0|\ll1\quad\text{for}\quad n\neq0,
\label{eqn:OptimizationII}
\end{equation}
purification is achieved quickly.
Equations (\ref{eqn:OptimizationI}) and
(\ref{eqn:OptimizationII}) are the conditions for the
\textit{optimal purification}, which we try to accomplish by
adjusting parameters $\tau$, $\ket{\phi}_\text{X}$, and those in
the Hamiltonian $H_\text{tot}$.

In the following sections, we discuss the above purification
scheme in more detail addressing a few specific examples, which
are so simple but still possess potential and practical
applications in quantum information and computation.

\section{Single-Qubit Purification}
\label{sec:Single}
Let us first observe how the above mechanism works in the
simplest example: we consider two qubits (two two-level systems)
X and A interacting with each other, whose total Hamiltonian is
given by
\begin{equation}
H_\text{tot}
=\Omega_\text{X}\frac{1+\sigma_3^\text{X}}{2}
+\Omega_\text{A}\frac{1+\sigma_3^\text{A}}{2}
+g(\sigma_+^\text{X}\sigma_-^\text{A}
+\sigma_-^\text{X}\sigma_+^\text{A}),
\label{eqn:HamiltonianSingle}
\end{equation}
where $\sigma_i\,(i=1,2,3)$ are the Pauli operators,
$\sigma_\pm=(\sigma_1\pm i\sigma_2)/2$ are the ladder operators,
and the frequencies $\Omega_\text{X(A)}$ and the coupling
constant $g\,(\neq0)$ are real parameters.
We repeatedly confirm the state of X and purify qubit A, i.e., we
discuss a purification of a single qubit.

The four eigenvalues of the total Hamiltonian $H_\text{tot}$ in
(\ref{eqn:HamiltonianSingle}) are given by
\begin{subequations}
\begin{align}
E^{(0)}&=0,\displaybreak[0]\\
E^{(1)}_\pm&=(\Omega_\text{X}+\Omega_\text{A})/2\pm\delta,
\displaybreak[0]\\
E^{(2)}&=\Omega_\text{X}+\Omega_\text{A},
\end{align}
\end{subequations}
and the corresponding eigenstates are
\begin{subequations}
\label{eqn:EigenstatesSingle}
\begin{align}
\ket{E^{(0)}}_\text{XA}
={}&\ket{\downarrow\downarrow}_\text{XA},\displaybreak[0]\\
\ket{E^{(1)}_\pm}_\text{XA}
={}&\frac{1}{\sqrt{2}}\,\biggl(
\epsilon(g)\sqrt{1\pm\frac{\Omega_\text{X}-\Omega_\text{A}}%
{2\delta}}\ket{\uparrow\downarrow}_\text{XA}
\nonumber\displaybreak[0]\\
&\phantom{\frac{1}{\sqrt{2}}\,\biggl(}
{}\pm\sqrt{1\mp\frac{\Omega_\text{X}-\Omega_\text{A}}{2\delta}}
\ket{\downarrow\uparrow}_\text{XA}
\biggr),\displaybreak[0]\\
\ket{E^{(2)}}_\text{XA}
={}&\ket{\uparrow\uparrow}_\text{XA},
\end{align}
\end{subequations}
where
\begin{equation}
\delta=\sqrt{(\Omega_\text{X}-\Omega_\text{A})^2/4+g^2},
\end{equation}
$\epsilon(g)$ is the sign function, and
$\ket{{\uparrow}({\downarrow})}$ is the eigenstate of the
operator $\sigma_3$ belonging to the eigenvalue $+1\,(-1)$ with
the phase convention $\ket{\uparrow}=\sigma_+\ket{\downarrow}$.
Hence, when the state of X, $\ket{\phi}_\text{X}$, is confirmed
repeatedly at time intervals $\tau$, the relevant operator to be
investigated, the projected time-evolution operator
$V_\phi(\tau)$, reads
\begin{widetext}
\begin{align}
V_\phi(\tau)
\equiv{}&
\bras{\phi}{\text{X}}e^{-iH_\text{tot}\tau}\ket{\phi}_\text{X}
\nonumber\displaybreak[0]\\
={}&\ketbras{\uparrow}{\uparrow}{\text{A}}
e^{-i(\Omega_\text{X}+\Omega_\text{A})\tau}\left[
\cos^2\!\frac{\theta}{2}
+e^{i(\Omega_\text{X}+\Omega_\text{A})\tau/2}\left(
\cos\delta\tau
+i\frac{\Omega_\text{X}-\Omega_\text{A}}{2\delta}\sin\delta\tau
\right)\sin^2\!\frac{\theta}{2}
\right]\nonumber\displaybreak[0]\\
&{}+\ketbras{\downarrow}{\downarrow}{\text{A}}\left[
\sin^2\!\frac{\theta}{2}
+e^{-i(\Omega_\text{X}+\Omega_\text{A})\tau/2}\left(
\cos\delta\tau
-i\frac{\Omega_\text{X}-\Omega_\text{A}}{2\delta}\sin\delta\tau
\right)\cos^2\!\frac{\theta}{2}
\right]\nonumber\displaybreak[0]\\
&{}-i\left(
\ketbras{\uparrow}{\downarrow}{\text{A}}e^{-i\varphi}
+\ketbras{\downarrow}{\uparrow}{\text{A}}e^{i\varphi}
\right)\frac{g}{\delta}
e^{-i(\Omega_\text{X}+\Omega_\text{A})\tau/2}
\sin\delta\tau\sin\frac{\theta}{2}\cos\frac{\theta}{2},
\label{eqn:Vsingle}
\end{align}
\end{widetext}
where the state $\ket{\phi}_\text{X}$ is parameterized as
\begin{equation}
\ket{\phi}_\text{X}
=e^{-i\varphi/2}\cos\frac{\theta}{2}\ket{\uparrow}_\text{X}
+e^{i\varphi/2}\sin\frac{\theta}{2}\ket{\downarrow}_\text{X}
\label{eqn:SpinParametrization}
\end{equation}
and the set of angles $(\theta,\varphi)$ characterizes the
``direction of `spin' X\@.''

If one of the two eigenvalues of the operator (\ref{eqn:Vsingle})
is larger in magnitude than the other, the condition for
purification (\ref{eqn:Condition}) is fulfilled, and qubit A is
purified into the eigenstate $\ket{u_0}_\text{A}$ belonging to
the larger (in magnitude) eigenvalue $\lambda_0$.
Furthermore, if condition (\ref{eqn:OptimizationI}),
$|\lambda_0|=1$, is satisfied, we can purify with a nonvanishing
success probability $P^{(\tau)}(N)\to\bras{\phi v_0}{\text{XA}}%
\varrho_\text{tot}\ket{\phi v_0}_\text{XA}$, and another
condition (\ref{eqn:OptimizationII}),
$|\lambda_1/\lambda_0|\ll1$, enables us to accomplish quick
purification.
We try to achieve these conditions by tuning the parameters.

The first adjustment for the optimal purification is
\begin{equation}
\theta=0\ \text{or}\ \pi,\quad\text{i.e.},\quad
\ket{\phi}_\text{X}=\ket{\uparrow}_\text{X}\ \text{or}\ %
\ket{\downarrow}_\text{X}
\label{eqn:OptimizationSingle}
\end{equation}
(see Appendix \ref{app:OptimalityProof}).
Actually, if we choose
$\ket{\phi}_\text{X}=\ket{\uparrow}_\text{X}$, the eigenvalues of
the projected time-evolution operator $V_\phi(\tau)$ are given by
\begin{subequations}
\begin{equation}
\begin{cases}
\lambda_0=e^{-i(\Omega_\text{X}+\Omega_\text{A})\tau},\\
\displaystyle
\lambda_1=e^{-i(\Omega_\text{X}+\Omega_\text{A})\tau/2}\left(
\cos\delta\tau-i\frac{\Omega_\text{X}-\Omega_\text{A}}{2\delta}
\sin\delta\tau
\right),
\end{cases}
\end{equation}
and the eigenvectors belonging to them are
\begin{equation}
\begin{cases}
\ket{u_0}_\text{A}=\ket{\uparrow}_\text{A},\\
\ket{u_1}_\text{A}=\ket{\downarrow}_\text{A},
\end{cases}
\quad
\begin{cases}
\bras{v_0}{\text{A}}=\bras{\uparrow}{\text{A}},\\
\bras{v_1}{\text{A}}=\bras{\downarrow}{\text{A}}.
\end{cases}
\end{equation}
\end{subequations}
It is clear that the magnitude of the eigenvalue $\lambda_0$ is
unity and that of $\lambda_1$,
\begin{equation}
|\lambda_1|
=\sqrt{1-\left(\frac{g}{\delta}\right)^2\sin^2\!\delta\tau},
\label{eqn:SecondEigenvalue}
\end{equation}
is less than unity provided
\begin{equation}
\delta\tau\neq n\pi\quad(n=1,2,\ldots).
\label{eqn:OptimizationSingleII}
\end{equation}
Both conditions (\ref{eqn:Condition}) and
(\ref{eqn:OptimizationI}) are thus satisfied, and according to
the theory presented in Sec.~\ref{sec:Framework}, we have an
optimal purification
\begin{equation}
\begin{cases}
\medskip
\varrho_\text{A}^{(\tau)}(N)
\to\ketbras{\uparrow}{\uparrow}{\text{A}},\\
P^{(\tau)}(N)
\to\bras{\uparrow\uparrow}{\text{XA}}\varrho_\text{tot}
\ket{\uparrow\uparrow}_\text{XA},
\end{cases}
\quad(N\to\infty).
\label{eqn:ToUp}
\end{equation}
After the repeated confirmations of the state
$\ket{\uparrow}_\text{X}$, qubit A is purified into
$\ket{\uparrow}_\text{A}$ with a \textit{nonvanishing}
probability $\bras{\uparrow\uparrow}{\text{XA}}\varrho_\text{tot}
\ket{\uparrow\uparrow}_\text{XA}$.
Similarly, another choice in (\ref{eqn:OptimizationSingle}),
i.e., a series of repeated confirmations of the state
$\ket{\downarrow}_\text{X}$, drives A into
$\ket{\downarrow}_\text{A}$ with a nonvanishing probability
$\bras{\downarrow\downarrow}{\text{XA}}\varrho_\text{tot}%
\ket{\downarrow\downarrow}_\text{XA}$:
\begin{equation}
\begin{cases}
\medskip
\varrho_\text{A}^{(\tau)}(N)
\to\ketbras{\downarrow}{\downarrow}{\text{A}},\\
P^{(\tau)}(N)
\to\bras{\downarrow\downarrow}{\text{XA}}\varrho_\text{tot}
\ket{\downarrow\downarrow}_\text{XA},
\end{cases}
\quad(N\to\infty).
\end{equation}
The final success probability $\bras{\uparrow\uparrow}{\text{XA}}
\varrho_\text{tot}\ket{\uparrow\uparrow}_\text{XA}$ for the
former choice $\ket{\phi}_\text{X}=\ket{\uparrow}_\text{X}$ or
$\bras{\downarrow\downarrow}{\text{XA}}\varrho_\text{tot}
\ket{\downarrow\downarrow}_\text{XA}$ for the latter
$\ket{\phi}_\text{X}=\ket{\downarrow}_\text{X}$ means that the
target state $\ket{\uparrow\uparrow}_\text{XA}$ or
$\ket{\downarrow\downarrow}_\text{XA}$ contained in the initial
state $\varrho_\text{tot}$ is fully extracted.
In this sense, the purification is optimal.

The second adjustment is for the fastest purification, which is
realized by the condition
\begin{equation}
\delta\tau=(n+1/2)\pi\quad(n=0,1,\ldots),
\label{eqn:FastestSingle}
\end{equation}
at which $|\lambda_1|$ in (\ref{eqn:SecondEigenvalue}) is the
smallest:
$|\lambda_1|=|\Omega_\text{X}-\Omega_\text{A}|/2\delta$.
We can achieve it by tuning the time interval $\tau$, for
instance.

\begin{figure}[b]
\includegraphics[width=0.45\textwidth]{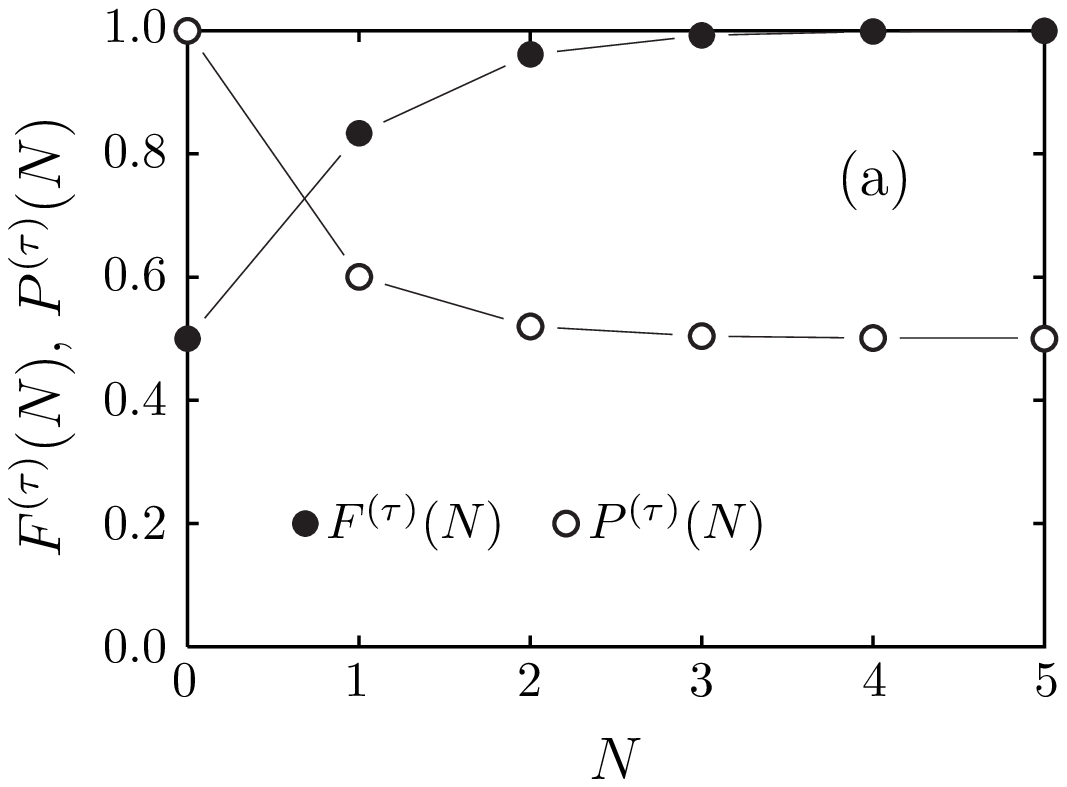}\\
\bigskip\smallskip
\includegraphics[width=0.45\textwidth]{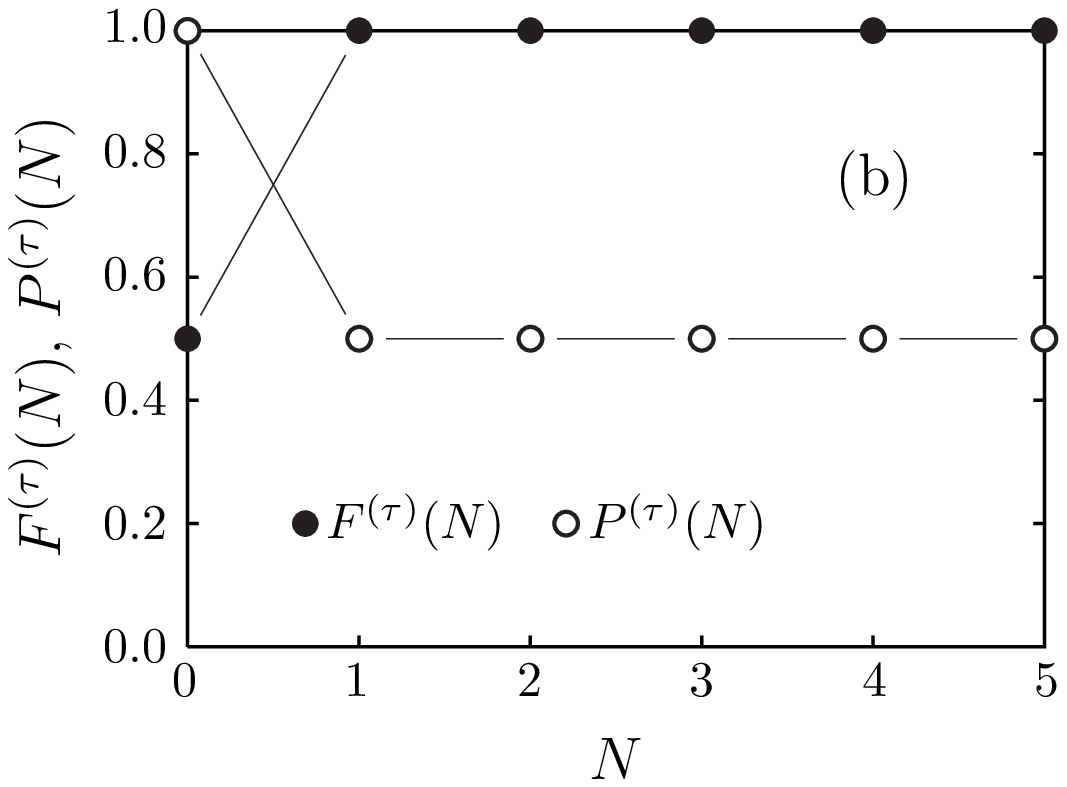}
\caption{Fidelity $F^{(\tau)}(N)$ and success probability
$P^{(\tau)}(N)$ for single-qubit purification.  The pure state
$\ket{{\uparrow}({\downarrow})}_\text{A}$ is extracted from the
initial mixed state $\varrho_\text{tot}
=\ketbras{{\uparrow}({\downarrow})}{{\uparrow}({\downarrow})}%
{\text{X}}\otimes(\ketbras{\uparrow}{\uparrow}{\text{A}}
+\ketbras{\downarrow}{\downarrow}{\text{A}})/2$ after repeated
confirmations of the state
$\ket{{\uparrow}({\downarrow})}_\text{X}$.  Parameters are
$\Omega_\text{X}=5$, $\Omega_\text{A}=6$,
$\tau=\pi/2\delta\simeq1.40$ for (a) and
$\Omega_\text{X}=\Omega_\text{A}$, $\tau=\pi/2\delta\simeq1.57$
for (b), in the unit such that $g=1$.  The time interval $\tau$
is tuned so as to satisfy the condition for the fastest
purification (\ref{eqn:FastestSingle}) in each case.}
\label{fig:FidelityYieldSingle}
\end{figure}
To be more explicit, let us demonstrate the extraction of the
pure state $\ket{\uparrow}_\text{A}$ from the initial mixed state
\begin{equation}
\varrho_\text{tot}=\ketbras{\uparrow}{\uparrow}{\text{X}}
\otimes\frac{1}{2}\left(\ketbras{\uparrow}{\uparrow}{\text{A}}
+\ketbras{\downarrow}{\downarrow}{\text{A}}\right).
\label{eqn:InitialStateSingle}
\end{equation}
After X is confirmed to be in the state $\ket{\uparrow}_\text{X}$
successfully $N$ times at time intervals $\tau$, the state of
qubit A and the probability for the successful confirmations read
\begin{equation}
\begin{cases}
\medskip
\displaystyle
\varrho_\text{A}^{(\tau)}(N)
=\frac{
\ketbras{\uparrow}{\uparrow}{\text{A}}
+[1-(g/\delta)^2\sin^2\!\delta\tau]^N
\ketbras{\downarrow}{\downarrow}{\text{A}}
}{1+[1-(g/\delta)^2\sin^2\!\delta\tau]^N},\\
\displaystyle
P^{(\tau)}(N)
=\frac{1}{2}\{1+[1-(g/\delta)^2\sin^2\!\delta\tau]^N\},
\end{cases}
\end{equation}
respectively, which clearly confirm the limits (\ref{eqn:ToUp})
unless $\delta\tau=n\pi$ ($n=1,2,\ldots$), and the convergences
are the fastest when the condition (\ref{eqn:FastestSingle}) is
satisfied.
(Note that $\bras{\uparrow\uparrow}{\text{XA}}\varrho_\text{tot}
\ket{\uparrow\uparrow}_\text{XA}=1/2$ for the initial state
considered here\@.)

In Fig.~\ref{fig:FidelityYieldSingle}(a), the success probability
$P^{(\tau)}(N)$ and the so-called fidelity to the target state
$\ket{u_0}_\text{A}=\ket{\uparrow}_\text{A}$, defined by
\begin{align}
F^{(\tau)}(N)
&\equiv\bras{u_0}{\text{A}}\varrho_\text{A}^{(\tau)}(N)
\ket{u_0}_\text{A}\nonumber\displaybreak[0]\\
&=\bras{\uparrow}{\text{A}}\varrho_\text{A}^{(\tau)}(N)
\ket{\uparrow}_\text{A},
\end{align}
are shown as functions of the number of measurements, $N$, for
the initial state (\ref{eqn:InitialStateSingle}), with the
parameters $\Omega_\text{X}=5$, $\Omega_\text{A}=6$, $g=1$,
$\tau=\pi/2\delta\simeq1.40$.
Since the condition (\ref{eqn:OptimizationI}), $|\lambda_0|=1$,
is fulfilled, the decay of the success probability
$P^{(\tau)}(N)$ is suppressed to yield the finite value
$\bras{\uparrow\uparrow}{\text{XA}}\varrho_\text{tot}
\ket{\uparrow\uparrow}_\text{XA}=1/2$, and since the time
interval $\tau$ is tuned so as to satisfy the condition for the
fastest purification (\ref{eqn:FastestSingle})
($|\lambda_1|\simeq0.45$), the pure state
$\ket{\uparrow}_\text{A}$ is extracted after only $N=4$ or $5$
measurements.
In an extreme case where $|\lambda_1|=0$ is possible, the
extraction is achieved just after one measurement.
Such a situation is depicted in
Fig.~\ref{fig:FidelityYieldSingle}(b) for the same initial state
as in Fig.~\ref{fig:FidelityYieldSingle}(a) with the parameter
set $\Omega_\text{X}=\Omega_\text{A}$, $g=1$,
$\tau=\pi/2\simeq1.57$.

\section{Initialization of Multiple Qubits}
\label{sec:Initialization}
The single-qubit purification in the previous section is too
simple but is easily extended for multi-qubit cases.
In the above example, one may realize that the state
$\ket{\uparrow\uparrow}_\text{XA}$ is an eigenstate of the total
Hamiltonian (\ref{eqn:HamiltonianSingle}) [see
(\ref{eqn:EigenstatesSingle})] and this is why the optimization
condition (\ref{eqn:OptimizationI}), $|\lambda_0|=1$, is achieved
with $\ket{\phi}_\text{X}=\ket{\uparrow}_\text{X}$ and
$\ket{u_0}_\text{A}=\ket{\uparrow}_\text{A}$ irrespectively of
the choice of the time interval $\tau$.
(The same argument applies to the case
$\ket{\phi}_\text{X}=\ket{\downarrow}_\text{X}$ there.)
\begin{figure}[b]
\includegraphics[width=0.45\textwidth]{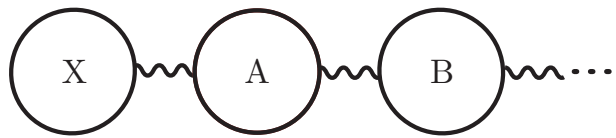}
\caption{A multi-qubit system with nearest-neighbor
interactions.}
\label{fig:MultiQubits}
\end{figure}
In the case of a multi-qubit system $\text{X}+\text{A}+\text{B}
+\cdots$ in Fig.~\ref{fig:MultiQubits}, with nearest-neighbor
interactions,
\begin{align}
H_\text{tot}
={}&\Omega_\text{X}\frac{1+\sigma_3^\text{X}}{2}
+\Omega_\text{A}\frac{1+\sigma_3^\text{A}}{2}
+\Omega_\text{B}\frac{1+\sigma_3^\text{B}}{2}
+\cdots\nonumber\displaybreak[0]\\
&{}+g_\text{XA}(\sigma_+^\text{X}\sigma_-^\text{A}
+\sigma_-^\text{X}\sigma_+^\text{A})
+g_\text{AB}(\sigma_+^\text{A}\sigma_-^\text{B}
+\sigma_-^\text{A}\sigma_+^\text{B})\nonumber\displaybreak[0]\\
&\phantom{{}+{}}{}+\cdots
\label{eqn:HamiltonianMultiple}
\end{align}
($g_\text{XA},g_\text{AB},\ldots\neq0$), the state
$\ket{{\uparrow\uparrow\uparrow}{\ldots}}_\text{XAB\ldots}$ is an
eigenstate of this total Hamiltonian $H_\text{tot}$, and it is
readily expected that the pure state
$\ket{{\uparrow\uparrow}{\ldots}}_\text{AB\ldots}$ is extracted
by repeated projections onto the state $\ket{\uparrow}_\text{X}$,
with the optimal success probability.
Similarly, repeated projections onto $\ket{\downarrow}_\text{X}$
set every qubit into $\ket{\downarrow}$ state, i.e., into
$\ket{{\downarrow\downarrow}{\ldots}}_\text{AB\ldots}$,
optimally.
This would be useful for \textit{initialization of multiple
qubits} in a quantum computer.

In order to make this idea more concrete, let us discuss in
detail with a three-qubit system $\text{X}+\text{A}+\text{B}$.
The important point is whether the condition for the purification
(\ref{eqn:Condition}) is achievable, i.e., whether all the
eigenvalues $\lambda_n$ except for the relevant one $\lambda_0$,
associated with the eigenstate $\ket{\uparrow\uparrow}_\text{AB}$
(or $\ket{\downarrow\downarrow}_\text{AB}$), can actually be less
than unity in magnitude.

For simplicity, we consider the case where
$\Omega_\text{X}=\Omega_\text{A}=\Omega_\text{B}=\Omega$.
The eight eigenvalues of the total Hamiltonian $H_\text{tot}$ are
given by
\begin{subequations}
\label{eqn:EigenvaluesMultiple}
\begin{align}
E^{(0)}&=0,\displaybreak[0]\\
E^{(1)}_0&=\Omega,&
E^{(1)}_\pm&=\Omega\pm\sqrt{2}\bar{g},\displaybreak[0]\\
E^{(2)}_0&=2\Omega,&
E^{(2)}_\pm&=2\Omega\pm\sqrt{2}\bar{g},\displaybreak[0]\\
E^{(3)}&=3\Omega,
\end{align}
\end{subequations}
and the corresponding eigenstates are
\begin{subequations}
\label{eqn:EigenstatesMultiple}
\begin{align}
\ket{E^{(0)}}_\text{XAB}
={}&\ket{\downarrow\downarrow\downarrow}_\text{XAB},
\displaybreak[0]\\
\ket{E^{(1)}_0}_\text{XAB}
={}&\cos\chi\ket{\downarrow\downarrow\uparrow}_\text{XAB}
-\sin\chi\ket{\uparrow\downarrow\downarrow}_\text{XAB},
\displaybreak[0]\\
\ket{E^{(1)}_\pm}_\text{XAB}
={}&\frac{1}{\sqrt{2}}(
\sin\chi\ket{\downarrow\downarrow\uparrow}_\text{XAB}
+\cos\chi\ket{\uparrow\downarrow\downarrow}_\text{XAB}
\nonumber\displaybreak[0]\\
&\phantom{\frac{1}{\sqrt{2}}(}
{}\pm\ket{\downarrow\uparrow\downarrow}_\text{XAB}
),\displaybreak[0]\\
\ket{E^{(2)}_0}_\text{XAB}
={}&\cos\chi\ket{\uparrow\uparrow\downarrow}_\text{XAB}
-\sin\chi\ket{\downarrow\uparrow\uparrow}_\text{XAB},
\displaybreak[0]\\
\ket{E^{(2)}_\pm}_\text{XAB}
={}&\frac{1}{\sqrt{2}}(
\sin\chi\ket{\uparrow\uparrow\downarrow}_\text{XAB}
+\cos\chi\ket{\downarrow\uparrow\uparrow}_\text{XAB}
\nonumber\displaybreak[0]\\
&\phantom{\frac{1}{\sqrt{2}}(}
{}\pm\ket{\uparrow\downarrow\uparrow}_\text{XAB}
),\displaybreak[0]\\
\ket{E^{(3)}}_\text{XAB}
={}&\ket{\uparrow\uparrow\uparrow}_\text{XAB},
\end{align}
\end{subequations}
where
\begin{gather}
\sqrt{2}\bar{g}=\sqrt{g_\text{XA}^2+g_\text{AB}^2},
\displaybreak[0]\\
\cos\chi=\frac{g_\text{XA}}{\sqrt{g_\text{XA}^2+g_\text{AB}^2}},
\quad
\sin\chi=\frac{g_\text{AB}}{\sqrt{g_\text{XA}^2+g_\text{AB}^2}}.
\end{gather}
Aiming at initializing qubits A and B into
$\ket{\downarrow\downarrow}_\text{AB}$, we repeatedly project X
onto the state $\ket{\downarrow}_\text{X}$ at time intervals
$\tau$, and the relevant operator to be investigated reads
\begin{align}
V_\downarrow(\tau)
\equiv{}&\bras{\downarrow}{\text{X}}e^{-iH_\text{tot}\tau}
\ket{\downarrow}_\text{X}\nonumber\displaybreak[0]\\
={}&\ketbras{\downarrow\downarrow}{\downarrow\downarrow}%
{\text{AB}}\nonumber\displaybreak[0]\\
&{}+\ketbras{\uparrow\downarrow}{\uparrow\downarrow}{\text{AB}}
e^{-i\Omega\tau}\cos\sqrt{2}\bar{g}\tau
\nonumber\displaybreak[0]\\
&{}+\ketbras{\downarrow\uparrow}{\downarrow\uparrow}{\text{AB}}
e^{-i\Omega\tau}(\cos^2\!\chi
+\sin^2\!\chi\cos\sqrt{2}\bar{g}\tau)
\nonumber\displaybreak[0]\\
&{}-i\ketbras{\uparrow\downarrow}{\downarrow\uparrow}{\text{AB}}
e^{-i\Omega\tau}\sin\chi\sin\sqrt{2}\bar{g}\tau
\nonumber\displaybreak[0]\\
&{}-i\ketbras{\downarrow\uparrow}{\uparrow\downarrow}{\text{AB}}
e^{-i\Omega\tau}\sin\chi\sin\sqrt{2}\bar{g}\tau
\nonumber\displaybreak[0]\\
&{}+\ketbras{\uparrow\uparrow}{\uparrow\uparrow}{\text{AB}}
e^{-2i\Omega\tau}(\sin^2\!\chi
+\cos^2\!\chi\cos\sqrt{2}\bar{g}\tau).
\end{align}
The target state $\ket{\downarrow\downarrow}_\text{AB}$ is an
eigenstate of this operator belonging to the eigenvalue
$\lambda_{\downarrow\downarrow}=1$, which satisfies the
optimization condition (\ref{eqn:OptimizationI}), and the other
three eigenvalues are give by
\begin{subequations}
\label{eqn:Initialize2Eigenvalues}
\begin{align}
\lambda_\pm
={}&e^{-i\Omega\tau}\biggl(
\cos^2\!\frac{\bar{g}\tau}{\sqrt{2}}
-\sin^2\!\chi\sin^2\!\frac{\bar{g}\tau}{\sqrt{2}}
\nonumber\displaybreak[0]\\
&{}\mp\sin\frac{\bar{g}\tau}{\sqrt{2}}
\sqrt{\cos^4\!\chi\sin^2\!\frac{\bar{g}\tau}{\sqrt{2}}
-4\sin^2\!\chi\cos^2\!\frac{\bar{g}\tau}{\sqrt{2}}}
\biggr),\displaybreak[0]\\
\lambda_{\uparrow\uparrow}
={}&e^{-2i\Omega\tau}\left(
1-2\cos^2\!\chi\sin^2\!\frac{\bar{g}\tau}{\sqrt{2}}
\right).
\label{eqn:Initialize2EigenvaluesUpUp}
\end{align}
\end{subequations}
If these three eigenvalues are all less than unity in magnitude,
the condition for the purification (\ref{eqn:Condition}) is
satisfied, and the initialized state
$\ket{\downarrow\downarrow}_\text{AB}$ is extracted from an
arbitrary mixed state $\varrho_\text{tot}$, with a nonvanishing
success probability $P^{(\tau)}(N)\to
\bras{\downarrow\downarrow\downarrow}{\text{XAB}}
\varrho_\text{tot}
\ket{\downarrow\downarrow\downarrow}_\text{XAB}$.
(Note that the left eigenvector belonging to the eigenvalue
$\lambda_{\downarrow\downarrow}$ is
$\bras{\downarrow\downarrow}{\text{AB}}$.)
Such a situation is realized provided
\begin{equation}
\sqrt{2}\bar{g}\tau\neq n\pi\quad(n=1,2,\ldots),
\label{eqn:InitializationCondition}
\end{equation}
which is clearly seen from Fig.~\ref{fig:EigenvaluesMultiple} and
a proof in Appendix \ref{app:EigenvaluesMultiple}.
\begin{figure}
\includegraphics[width=0.45\textwidth]{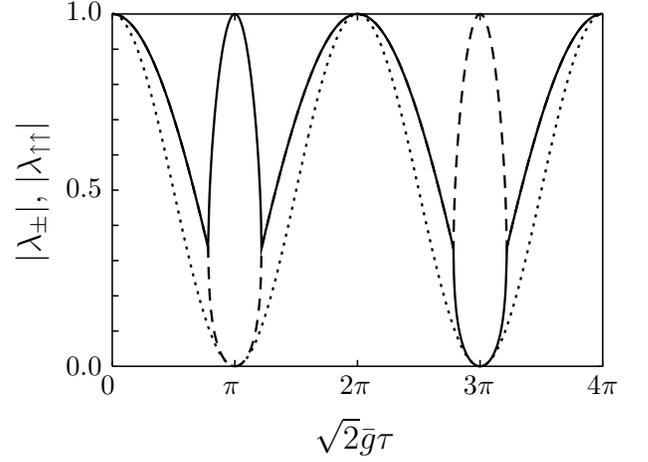}
\caption{Magnitudes of the eigenvalues $\lambda_+$ (solid line),
$\lambda_-$ (dashed line), and $\lambda_{\uparrow\uparrow}$
(dotted line) in (\ref{eqn:Initialize2Eigenvalues}), as functions
of $\sqrt{2}\bar{g}\tau$.
In this figure, we set $g_\text{XA}=g_\text{AB}$.
Note that $|\lambda_+|=|\lambda_-|$ within each range
$2n\pi-\zeta\le\sqrt{2}\bar{g}\tau\le2n\pi+\zeta$
($n=0,1,\ldots$), where $\zeta$ ($0<\zeta<\pi$) is defined by
$\tan(\zeta/2)=2|\sin\chi|/\cos^2\!\chi$, and
$\zeta\simeq0.78\pi$ when $g_\text{XA}=g_\text{AB}$.}
\label{fig:EigenvaluesMultiple}
\end{figure}
\begin{figure}
\includegraphics[width=0.45\textwidth]{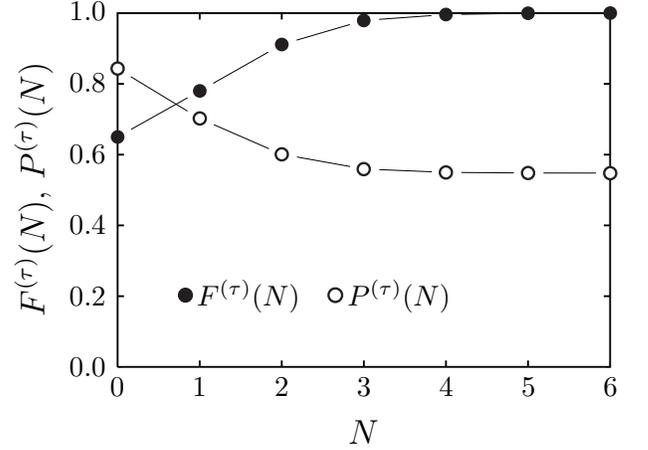}
\caption{Fidelity $F^{(\tau)}(N)$ and success probability
$P^{(\tau)}(N)$ for two-qubit initialization.  Through the
repeated confirmations of the state $\ket{\downarrow}_\text{X}$,
qubits A and B are initialized into
$\ket{\downarrow\downarrow}_\text{AB}$ from the thermal
equilibrium state of the total system at temperature $T$, i.e.,
$\varrho_\text{tot}\propto e^{-\beta H_\text{tot}}$ with
$\beta=(k_\text{B}T)^{-1}$.
Parameters are $g_\text{XA}=g_\text{AB}=1$, $\Omega=2$,
$\tau=\zeta/\sqrt{2}\simeq1.73$, $k_\text{B}T=\beta^{-1}=1$.  The
time interval $\tau$ is tuned so as to make
$\max(|\lambda_+|,|\lambda_-|,|\lambda_{\uparrow\uparrow}|)$ the
smallest, which is for the fastest initialization (see
Fig.~\ref{fig:EigenvaluesMultiple}).}
\label{fig:Initialize2}
\end{figure}
The final success probability $P^{(\tau)}(N)\to
\bras{\downarrow\downarrow\downarrow}{\text{XAB}}
\varrho_\text{tot}
\ket{\downarrow\downarrow\downarrow}_\text{XAB}$ is again
optimal, in the sense that the target state
$\ket{\downarrow\downarrow\downarrow}_\text{XAB}$ contained in
the initial state $\varrho_\text{tot}$ is fully extracted.

The above argument reveals the possibility of initialization at
least for two qubits.
Initialization of two qubits into
$\ket{\downarrow\downarrow}_\text{AB}$ from the thermal
equilibrium state of the total system at temperature $T$, i.e.,
$\varrho_\text{tot}\propto e^{-\beta H_\text{tot}}$ with
$\beta=(k_\text{B}T)^{-1}$, is demonstrated in
Fig.~\ref{fig:Initialize2}.
Note that it is effective when $\Omega>\sqrt{2}\bar{g}$, since in
such a case, $\ket{\downarrow\downarrow\downarrow}_\text{XAB}$ is
the ground state of the total system.
The analytic formula for the final success probability is
$P^{(\tau)}(\infty)=[1+(e^{-\beta\Omega}+e^{-2\beta\Omega})(1
+2\cosh\sqrt{2}\beta\bar{g})+e^{-3\beta\Omega}]^{-1}$.

It is natural to expect that the same mechanism also works for
systems with more qubits as in Fig.~\ref{fig:MultiQubits}.
It is hard to imagine that the magnitudes of eigenvalues of
$V_\downarrow(\tau)$ other than the relevant one
$\lambda_{\downarrow\downarrow\ldots}$ (whose magnitude is unity)
is also unity \textit{irrespectively of the values of
parameters}.
Further detailed investigations on its efficiency, robustness,
and so on, will certainly clarify the possibility of a new useful
procedure for initializing multiple qubits.

\section{Entanglement Purification}
\label{sec:EntanglementPurification}
\begin{figure}
\includegraphics[width=0.3\textwidth]{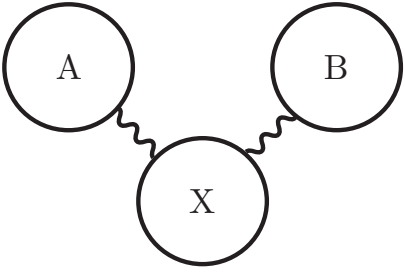}
\caption{We repeat measurements on qubit X and extract one of the
Bell states, $\ket{\Psi^-}_\text{AB}\equiv
(\ket{\uparrow\downarrow}_\text{AB}
-\ket{\downarrow\uparrow}_\text{AB})/\sqrt{2}$, in qubits A and
B\@.}
\label{fig:ThreeQubits}
\end{figure}

One of the most significant issues in the field of quantum
information and computation is how to prepare
\textit{entanglement}, and therefore, it is interesting and
important to examine whether the present scheme can realize
\textit{entanglement purification/preparation}.
We show, in this section, that it is actually possible.
In order to demonstrate it explicitly, let us discuss a simple
Hamiltonian
\begin{align}
H_\text{tot}
={}&\Omega\frac{1+\sigma_3^\text{X}}{2}
+\Omega\frac{1+\sigma_3^\text{A}}{2}
+\Omega\frac{1+\sigma_3^\text{B}}{2}\nonumber\displaybreak[0]\\
&{}+g(\sigma_+^\text{X}\sigma_-^\text{A}
+\sigma_-^\text{X}\sigma_+^\text{A})
+g(\sigma_+^\text{X}\sigma_-^\text{B}
+\sigma_-^\text{X}\sigma_+^\text{B}).
\label{eqn:HamiltonianEntanglement}
\end{align}
The control qubit X is coupled to qubits A and B as in
Fig.~\ref{fig:ThreeQubits}.
We confirm X to be in the state $\ket{\phi}_\text{X}$ repeatedly
at time intervals $\tau$ and end up with an extraction of an
entanglement between A and B, which are initially in a mixed
state $\varrho_\text{tot}$.

The spectrum of the total Hamiltonian $H_\text{tot}$ is already
given in (\ref{eqn:EigenvaluesMultiple}) with $\bar{g}$ replaced
by $g$, and the eigenstates are
\begin{subequations}
\label{eqn:EigenstatesEntanglement}
\begin{align}
\ket{E^{(0)}}_\text{XAB}
={}&\ket{\downarrow\downarrow\downarrow}_\text{XAB},
\displaybreak[0]\\
\ket{E^{(1)}_0}_\text{XAB}
={}&\ket{{\downarrow}\Psi^-}_\text{XAB},\displaybreak[0]\\
\ket{E^{(1)}_\pm}_\text{XAB}
={}&\frac{1}{\sqrt{2}}[\ket{{\downarrow}\Psi^+}_\text{XAB}
\pm\epsilon(g)\ket{\uparrow\downarrow\downarrow}_\text{XAB}],
\displaybreak[0]\\
\ket{E^{(2)}_0}_\text{XAB}
={}&\ket{{\uparrow}\Psi^-}_\text{XAB},\displaybreak[0]\\
\ket{E^{(2)}_\pm}_\text{XAB}
={}&\frac{1}{\sqrt{2}}[\ket{{\uparrow}\Psi^+}_\text{XAB}
\pm\epsilon(g)\ket{\downarrow\uparrow\uparrow}_\text{XAB}],
\displaybreak[0]\\
\ket{E^{(3)}}_\text{XAB}
={}&\ket{\uparrow\uparrow\uparrow}_\text{XAB}.
\end{align}
\end{subequations}
The relevant projected time-evolution operator is given, in this
case, by
\begin{widetext}
\begin{align}
V_\phi(\tau)
\equiv{}&\bras{\phi}{\text{X}}e^{-iH_\text{tot}\tau}
\ket{\phi}_\text{X}\nonumber\displaybreak[0]\\
={}&\ketbras{\Psi^-}{\Psi^-}{\mathrm{AB}}
e^{-i\Omega\tau}\left(
\sin^2\!\frac{\theta}{2}+e^{-i\Omega\tau}\cos^2\!\frac{\theta}{2}
\right)\nonumber\displaybreak[0]\\
&{}+\ketbras{\downarrow\downarrow}{\downarrow\downarrow}%
{\mathrm{AB}}\left(
\sin^2\!\frac{\theta}{2}
+e^{-i\Omega\tau}\cos\sqrt{2}g\tau\cos^2\!\frac{\theta}{2}
\right)\nonumber\displaybreak[0]\\
&{}+\ketbras{\Psi^+}{\Psi^+}{\mathrm{AB}}
e^{-i\Omega\tau}\cos\sqrt{2}g\tau\left(
\sin^2\!\frac{\theta}{2}+e^{-i\Omega\tau}\cos^2\!\frac{\theta}{2}
\right)\nonumber\displaybreak[0]\\
&{}+\ketbras{\uparrow\uparrow}{\uparrow\uparrow}{\mathrm{AB}}
e^{-2i\Omega\tau}\left(
\cos\sqrt{2}g\tau\sin^2\!\frac{\theta}{2}
+e^{-i\Omega\tau}\cos^2\!\frac{\theta}{2}
\right)\nonumber\displaybreak[0]\\
&{}-i\left(
\ketbras{\downarrow\downarrow}{\Psi^+}{\mathrm{AB}}e^{i\varphi}
+\ketbras{\Psi^+}{\downarrow\downarrow}{\mathrm{AB}}e^{-i\varphi}
\right)e^{-i\Omega\tau}\sin\sqrt{2}g\tau\sin\frac{\theta}{2}
\cos\frac{\theta}{2}\nonumber\displaybreak[0]\\
&{}-i\left(
\ketbras{\uparrow\uparrow}{\Psi^+}{\mathrm{AB}}e^{-i\varphi}
+\ketbras{\Psi^+}{\uparrow\uparrow}{\mathrm{AB}}e^{i\varphi}
\right)e^{-2i\Omega\tau}\sin\sqrt{2}g\tau\sin\frac{\theta}{2}
\cos\frac{\theta}{2},
\end{align}
\end{widetext}
where $\ket{\Psi^\pm}_\text{AB}$ are the two of the four Bell
states $\ket{\Psi^\pm}_\text{AB}
=(\ket{\uparrow\downarrow}_\text{AB}
\pm\ket{\downarrow\uparrow}_\text{AB})/\sqrt{2}$,
$\ket{\Phi^\pm}_\text{AB}=(\ket{\uparrow\uparrow}_\text{AB}
\pm\ket{\downarrow\downarrow}_\text{AB})/\sqrt{2}$, and
$\ket{\phi}_\text{X}$ is parameterized as in
(\ref{eqn:SpinParametrization}).
Since the Hamiltonian (\ref{eqn:HamiltonianEntanglement}) is
symmetric under the exchange between A and B, $V_\phi(\tau)$
splits into two sectors: the singlet sector and the triplet one.
The singlet state $\ket{\Psi^-}_\mathrm{AB}$ is apparently one of
the four eigenstates of $V_\phi(\tau)$ belonging to the
eigenvalue
\begin{equation}
\lambda_{\Psi^-}=e^{-i\Omega\tau}\left(
\sin^2\!\frac{\theta}{2}+e^{-i\Omega\tau}\cos^2\!\frac{\theta}{2}
\right),
\end{equation}
and hence, we can extract an entangled state, i.e., the Bell
state $\ket{\Psi^-}_\text{AB}$, after a number of measurements on
X, provided (i) the eigenvalue $\lambda_{\Psi^-}$ is larger in
magnitude than any other eigenvalues.
Furthermore, if (ii) condition (\ref{eqn:OptimizationI}), i.e.,
$|\lambda_{\Psi^-}|=1$, is achieved, $\ket{\Psi^-}_\text{AB}$ is
extracted with an optimal probability $P^{(\tau)}(N)\to
\bras{\phi\Psi^-}{\text{XAB}}\varrho_\text{tot}
\ket{\phi\Psi^-}_\text{XAB}$, which is again optimal in the same
sense as in the preceding examples, i.e., the target entangled state $\ket{\Psi^-}_\text{AB}$ contained in the initial state $\varrho_\text{tot}$ has been fully extracted.

Requirement (ii) is fulfilled by the choice of the parameters as
$|\Omega|\tau=2n\pi$ ($n=0,1,\ldots$) or $\sin\theta=0$, but the
latter choice violates requirement (i).
It is, therefore, necessary that
\begin{subequations}
\label{eqn:ConditionEntanglement}
\begin{equation}
|\Omega|\tau=2n\pi\ (n=0,1,\ldots)\quad\text{and}\quad
\ket{\phi}_\text{X}\neq\ket{\uparrow}_\text{X},
\ket{\downarrow}_\text{X}.
\label{eqn:ConditionEntanglementI}
\end{equation}
(Note that the first condition is automatically satisfied without
tuning the time interval $\tau$, when $\Omega=0$.)
Furthermore, one can prove as in Appendix
\ref{app:EigenvaluesEntanglement} that requirement (i) is met,
under the condition (\ref{eqn:ConditionEntanglementI}), provided
\begin{equation}
|g|\tau/\sqrt{2}\neq m\pi/2\quad(m=1,2,\ldots).
\label{eqn:ConditionEntanglementII}
\end{equation}
\end{subequations}
The existence of such a parameter set satisfying
(\ref{eqn:ConditionEntanglement}) explicitly discloses the
possibility of extracting entanglement through Zeno-like
measurements.

In the case of the choice
\begin{equation}
\ket{\phi}_\text{X}
=\ket{\rightarrow}_\text{X}
\equiv\frac{1}{\sqrt{2}}\left(
\ket{\uparrow}_\text{X}+\ket{\downarrow}_\text{X}
\right),
\end{equation}
for example, the four eigenvalues are given by
\begin{subequations}
\begin{align}
\lambda_{\Psi^-}&=1,\quad
\lambda_{\Phi^-}=\cos^2\!\frac{g\tau}{\sqrt{2}},
\displaybreak[0]\\
\lambda_\pm&=1-\frac{1}{2}\sin\frac{g\tau}{\sqrt{2}}\left(
3\sin\frac{g\tau}{\sqrt{2}}
\pm\epsilon(g)\sqrt{1-9\cos^2\!\frac{g\tau}{\sqrt{2}}}
\right),
\end{align}
\end{subequations}
whose magnitudes behave as in Fig.~\ref{fig:EigenvaluesMultiple}
but with $\lambda_{\uparrow\uparrow}$ replaced by
$\lambda_{\Phi^-}$.
\begin{figure}[t]
\includegraphics[width=0.45\textwidth]{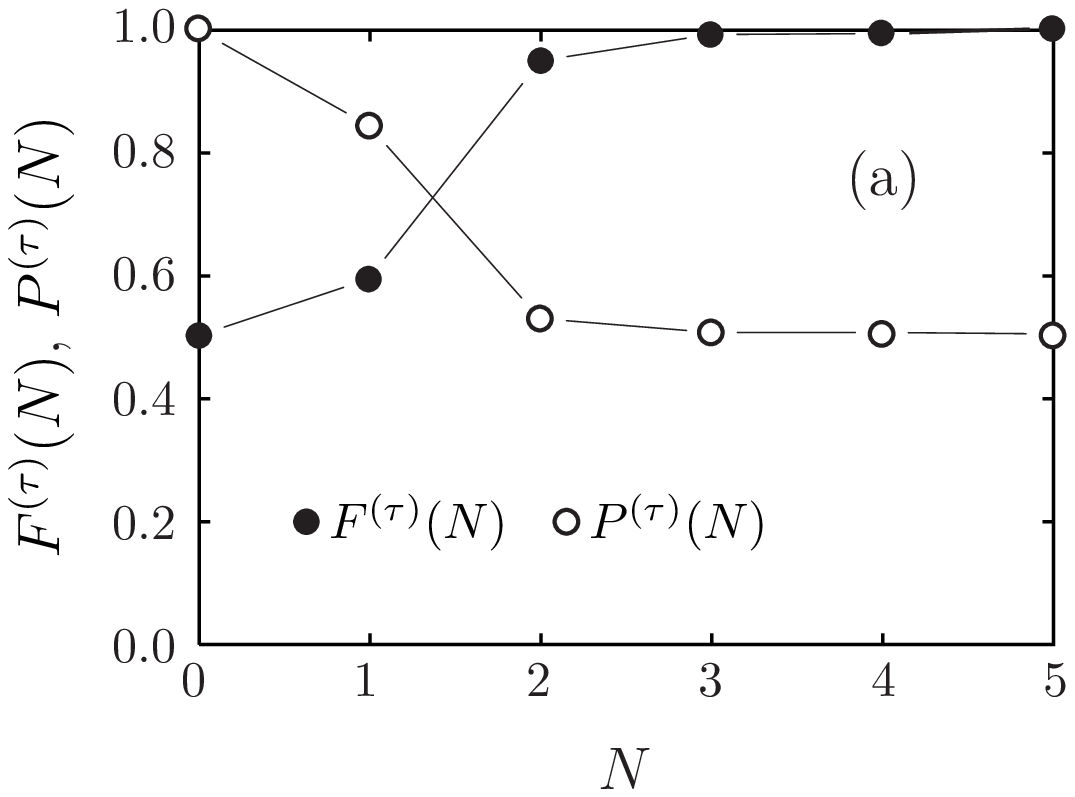}\\
\bigskip\smallskip
\includegraphics[width=0.45\textwidth]{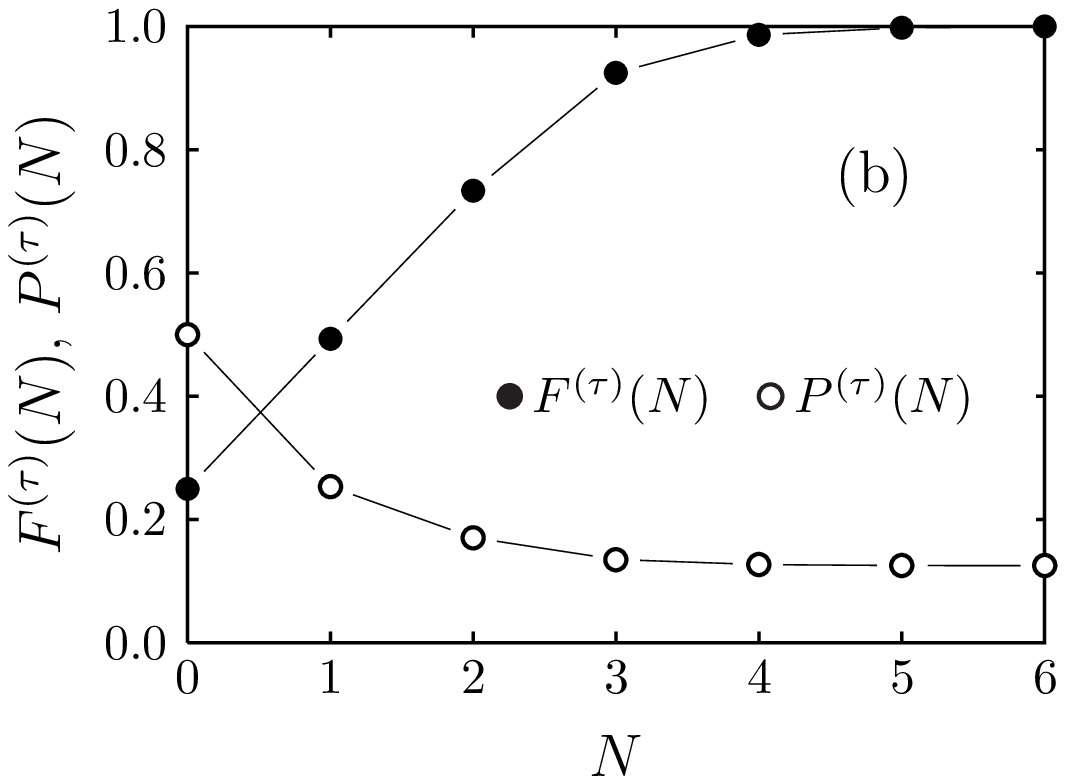}
\caption{Fidelity $F^{(\tau)}(N)$ and success probability
$P^{(\tau)}(N)$ for entanglement purification.  The entangled
state $\ket{\Psi^-}_\text{AB}$ is extracted from (a) a product
state $\varrho_\text{tot}=\ketbras{\rightarrow}{\rightarrow}%
{\text{X}}\otimes\ketbras{\uparrow}{\uparrow}{\text{A}}\otimes
\ketbras{\downarrow}{\downarrow}{\text{B}}$ and (b) the thermal
state $\varrho_\text{tot}\propto e^{-\beta H_\text{tot}}$ at
temperature $T=(k_\text{B}\beta)^{-1}$, through repeated
confirmations of the state $\ket{\rightarrow}_\text{X}$.
Parameters are $\Omega=0$, $\tau=0.5\pi\simeq1.57$ for (a), and
$\Omega=0$, $\tau=\zeta/\sqrt{2}\simeq1.73$,
$k_\text{B}T=\beta^{-1}=\infty$ for (b), in the unit such that
$g=1$, where $\zeta$ is defined in the caption of
Fig.~\ref{fig:EigenvaluesMultiple}.  For the initial thermal
state in (b) with $\Omega=0$, the success probability for the
zeroth confirmation is given by $P^{(\tau)}(0)=1/2$ for any set
of parameters $(\theta, \varphi, g, T)$, and the final value
$P^{(\tau)}(\infty)=[8\cosh^2(\beta g/\sqrt{2})]^{-1}$ becomes
largest at $k_\text{B}T/|g|=(\beta|g|)^{-1}=\infty$.}
\label{fig:Entanglement}
\end{figure}

The extraction of the entangled state $\ket{\Psi^-}_\text{AB}$ is
dem\-on\-strated in Fig.~\ref{fig:Entanglement}, from a product
state $\varrho_\text{tot}=\ketbras{\rightarrow}{\rightarrow}%
{\text{X}}\otimes\ketbras{\uparrow}{\uparrow}{\text{A}}\otimes
\ketbras{\downarrow}{\downarrow}{\text{B}}$ and from the thermal
equilibrium state $\varrho_\text{tot}\propto
e^{-\beta H_\text{tot}}$ at temperature
$T=(k_\text{B}\beta)^{-1}$.

\section{Concluding Remarks}
\label{sec:Summary}
The examples presented in this article demonstrate how the present purification scheme works, and suggest a few potential applications, even though the analyses are heuristically based and no general ``optimization'' theory or strategy has been given.
Remarkable features of the scheme are summarized as follows.
(i) The first point is the simplicity.
Many of the other proposed procedures are composed of several
steps with different operations, such as rotation, \textsc{cnot}
operation, and measurement \cite{ref:QuantInfoCompZeilinger,%
ref:PurificationBennett}.
In the present scheme, on the other hand, one has only to repeat
one and the same measurement.
(ii) Furthermore, the ``optimal'' success probability is possible
in the sense that the target state contained in the initial state
is fully extracted.
In several other methods \cite{ref:QuantInfoCompZeilinger,%
ref:PurificationBennett}, on the contrary, it decays to zero as
the fidelity approaches unity \cite{note:Purification}.
(iii) The number of measurements required for purification is
considerably reduced by appropriate choices of parameters, and
purification is attainable after only a few steps.

Another point to be stressed is the flexibility.
While many of the other schemes
\cite{ref:QuantInfoCompZeilinger,ref:PurificationBennett} are
designed for specific systems, the framework is presented in
Sec.~\ref{sec:Framework} on a general setting, and there are
diverse systems and purposes which fit the present scheme.
We have already observed, in this article, two different
applications on the same idea: initialization and entanglement
purification.
Additional ideas or slight modifications to the basic scheme
would provide us with various methods of state preparation.
An interesting extension of the present scheme is for extraction
of entanglement \textit{between two spatially-separated qubits}
\cite{ref:QuantInfoCompZeilinger,ref:EntanglementSeparate,%
ref:PurificationBennett,ref:PurificationExperiments}, which is
often necessary for quantum communication, quantum teleportation,
and so on.
(The original protocols for entanglement purification
\cite{ref:QuantInfoCompZeilinger,ref:PurificationBennett} are
aimed at this purpose.)
It is actually possible and will be reported elsewhere \cite{ref:qpfes}.
One of the other possible extensions is to go beyond a method of extracting quantum state.
It would be interesting, for example, if we could find a novel method of \textit{transferring} quantum state \cite{ref:StateTransfer,ref:StateTransferRev} rather than extracting it.

In this article, only qubit systems, i.e., finite-di\-men\-sional
systems, have been discussed.
One has to keep in mind that the condition (\ref{eqn:Condition}) plays a crucial role in the present purification scheme.
If this condition is met, however, it works for infinite-dimensional ones as well.
In fact, a harmonic oscillator, which has an infinite number of
energy levels, can be purified through the present method, which
is explicitly demonstrated in \cite{ref:qpf}.
This also shows the broad range of applicability of the scheme.
It is not obvious, however, whether one can purify systems with
continuous spectra, since they seem, at first sight, unlikely to
satisfy the condition for purification (\ref{eqn:Condition}),
especially the discreteness of the eigenvalues.
This point is one of the interesting future subjects, since it
would be required in some cases to purify quantum states in the
presence of environmental systems, namely, under dissipation
and/or dephasing.

The simplicity and the efficiency mentioned above would
facilitate practical experimental applications of the present
scheme.
The flexibility allows one to apply it to various kinds of
systems intended for quantum information and computation, such as
optical setups \cite{ref:PurificationExperiments}, ion-trap
systems \cite{ref:NIST,ref:Blatt}, solid-state quantum computers
\cite{ref:NEC}, and so on.
In practice, one should face many unwanted factors, and
robustness of the method against them is crucial.
In the present scheme, it is often required to tune certain
parameters in order to extract a desired pure state, and it is an
important subject to clarify how precise the tuning should be and
how much error the method suffers from when the parameters are
mistuned.
It is also a remained issue to explore how ideal projective
measurements are realized in actual experiments.
Investigations on these points are now in progress.

\begin{acknowledgments}
The authors acknowledge useful and helpful discussions with
Professor I. Ohba.
This work is partly supported by a Grant for The 21st Century COE
Program (Physics of Self-Organization Systems) at Waseda
University and a Grant-in-Aid for Priority Areas Research (B)
(No.~13135221) from the Ministry of Education, Culture, Sports,
Science and Technology, Japan, by a Grant-in-Aid for Scientific
Research (C) (No.~14540280) from the Japan Society for the
Promotion of Science, by a Waseda University Grant for Special
Research Projects (No.~2002A-567), and by the bilateral
Italian-Japanese project 15C1 on ``Quantum Information and
Computation'' of the Italian Ministry for Foreign Affairs.
\end{acknowledgments}

\appendix
\section{Bound on the Eigenvalues of $V_\phi(\tau)$}
\label{app:Bound}
Let us prove that the eigenvalues $\lambda_n$ of the projected
time-evolution operator $V_\phi(\tau)$ are bounded as in
(\ref{eqn:Bound}).

For an arbitrary state of A, say $\ket{\psi}_\text{A}$,
\begin{align}
0&\le\|V_\phi(\tau)\ket{\psi}_\text{A}\|^2
\nonumber\displaybreak[0]\\
&=\|\bras{\phi}{\text{X}}e^{-iH_\text{tot}\tau}
\ket{\phi}_\text{X}\ket{\psi}_\text{A}\|^2
\nonumber\displaybreak[0]\\
&\le\|e^{-iH_\text{tot}\tau}\ket{\phi}_\text{X}
\ket{\psi}_\text{A}\|^2
\nonumber\displaybreak[0]\\
&=1.
\label{eqn:Inequality}
\end{align}
Hence, by setting $\ket{\psi}_\text{A}=\ket{u_n}_\text{A}$ [a
right eigenvector of the operator $V_\phi(\tau)$] and noting
$\|V_\phi(\tau)\ket{u_n}_\text{A}\|^2=|\lambda_n|^2$, we obtain
the inequality (\ref{eqn:Bound}).
As is clear from this proof, the bound (\ref{eqn:Bound}) reflects
unitarity of the time-evolution operator
$e^{-iH_\text{tot}\tau}$.

\section{A Nondiagonalizable $V_\phi(\tau)$ Case}
\label{app:JordanDecomposition}
It is assumed in Sec.~\ref{sec:Framework} that the projected
time-evolution operator $V_\phi(\tau)$ is diagonalized like
(\ref{eqn:SpectralDecomp}), but it is not the case if some of its
eigenvalues are degenerated.
Here we show, however, that the assumption of the
diagonalizability is not essential to the purification.

When an eigenvalue $\lambda_n$ of the (finite-dimensional) operator $V_\phi(\tau)$ is $M_n$-fold degenerate, there do not always exist $M_n$ linearly independent eigenvectors.
This fact spoils the diagonalizability of the operator $V_\phi(\tau)$.
There exist $d_n\,(\le M_n)$ linearly independent right eigenvectors $\ket{u_n^{(k)}}_\text{A}$ ($k=1,\ldots,d_n$) belonging to the eigenvalue $\lambda_n$ ($d_n$ is called ``dimension of the eigenspace''), and one can find $M_n-d_n$ linearly independent ``generalized eigenvectors'' $\ket{u_n^{(k)}}_\text{A}$ ($k=d_n+1,\ldots,M_n$) which are subjected to the conditions
\begin{multline}
V_\phi(\tau)\ket{u_n^{(k)}}_\text{A}
=\lambda_n\ket{u_n^{(k)}}_\text{A}+\ket{u_n^{(k-1)}}_\text{A}\\
(k=d_n+1,\ldots,M_n)
\end{multline}
and linearly independent of the eigenvectors $\ket{u_n^{(k)}}_\text{A}$ ($k=1,\ldots,d_n$) \cite{ref:Kato}.
The right vectors $\ket{u_n^{(k)}}_\text{A}$ ($k=1,\ldots,M_n$)
then form a complete set within the subspace associated with the
eigenvalue $\lambda_n$, and there exist corresponding left
vectors $\bras{v_n^{(k)}}{\text{A}}$ ($k=1,\ldots,M_n$), which
satisfy the orthonormality
\begin{subequations}
\begin{equation}
\brackets{v_m^{(k)}}{u_n^{(\ell)}}{\text{A}}
=\delta_{mn}\delta_{k\ell}
\end{equation}
and completeness conditions
\begin{equation}
\sum_n\mathcal{P}_n=\openone_\text{A},\quad
\mathcal{P}_n
=\sum_{k=1}^{M_n}\ketbras{u_n^{(k)}}{v_n^{(k)}}{\text{A}}.
\end{equation}
\end{subequations}
The operator $V_\phi(\tau)$ is now expanded as
\begin{subequations}
\label{eqn:Jordan}
\begin{equation}
V_\phi(\tau)=\sum_n(\lambda_n\mathcal{P}_n+\mathcal{D}_n)
\end{equation}
with
\begin{equation}
\mathcal{D}_n
=\sum_{k=d_n+1}^{M_n}\ketbras{u_n^{(k-1)}}{v_n^{(k)}}{\text{A}},
\end{equation}
\end{subequations}
which is the most general form of spectral decomposition and is called ``Jordan canonical form'' \cite{ref:Kato}.
Note the relations
\begin{subequations}
\begin{align}
\mathcal{P}_m\mathcal{P}_n&=\delta_{mn}\mathcal{P}_n,
\displaybreak[0]\\
\mathcal{D}_m\mathcal{P}_n&=\mathcal{P}_n\mathcal{D}_m
=\mathcal{D}_m\delta_{mn},\displaybreak[0]\\
\mathcal{D}_m\mathcal{D}_n&=0\quad(m\neq n),
\end{align}
and
\begin{equation}
\mathcal{D}_n^{M_n-d_n+1}=0.
\end{equation}
\end{subequations}

From the spectral decomposition (\ref{eqn:Jordan}), it is easily
deduced that
\begin{subequations}
\begin{equation}
\bm{(}V_\phi(\tau)\bm{)}^N
=\sum_n\biggl(
\lambda_n^N\mathcal{P}_n
+\!\!\sum_{r=1}^{\min(N,M_n-d_n)}\!\!\!\!
{}_NC_r\,\lambda_n^{N-r}\mathcal{D}_n^r
\biggr),
\label{eqn:GeneralizedExpansion}
\end{equation}
where
\begin{equation}
\mathcal{D}_n^r
=\sum_{k=d_n+r}^{M_n}\ketbras{u_n^{(k-r)}}{v_n^{(k)}}{\text{A}}.
\end{equation}
\end{subequations}
Therefore, \textit{if the largest (in magnitude) eigenvalue  is
unique}, which is denoted by $\lambda_0$, \textit{and
nondegenerate} (i.e., $M_0=1$, $d_0=1$, $D_0=0$), the single term
in the expansion (\ref{eqn:GeneralizedExpansion}) again dominates
asymptotically like (\ref{eqn:Mechanism}) (note that ${}_NC_r\sim
N^r/r!$ for large $N$), which leads to the same conclusion as
(\ref{eqn:Purification}).
The purification does not suffer from degeneracy in the other
eigenvalues than the largest (in magnitude) one $\lambda_0$.
The crucial condition to the purification is
(\ref{eqn:Condition}).

\section{Optimization of the Single-Qubit Purification}
\label{app:OptimalityProof}
We show here that the condition (\ref{eqn:OptimizationSingle})
together with (\ref{eqn:OptimizationSingleII}) is the necessary
and sufficient condition for the optimal purification with both
(\ref{eqn:Condition}) and (\ref{eqn:OptimizationI}) for model
(\ref{eqn:HamiltonianSingle}).

First, we try to achieve the upper bound in the inequality
(\ref{eqn:Inequality}), i.e.,
$\|V_\phi(\tau)\ket{\psi}_\text{A}\|=1$, in model
(\ref{eqn:HamiltonianSingle}).
If such a state $\ket{\psi}_\text{A}$ is found and is an
eigenstate of the operator $V_\phi(\tau)$, say
$\ket{u_n}_\text{A}$, we have $|\lambda_n|=1$.
As is easily seen from (\ref{eqn:Inequality}), the equality holds
only when
\begin{equation}
\bras{\phi_\perp}{\text{X}}e^{-iH_\text{tot}\tau}
\ket{\phi}_\text{X}\ket{\psi}_\text{A}=0
\label{eqn:EqualityCondition}
\end{equation}
is satisfied, where $\ket{\phi_\perp}_\text{X}$ is a vector
perpendicular to $\ket{\phi}_\text{X}$ in
(\ref{eqn:SpinParametrization}), i.e.,
\begin{equation}
\ket{\phi_\perp}_\text{X}
=e^{-i\varphi/2}\sin\frac{\theta}{2}\ket{\uparrow}_\text{X}
-e^{i\varphi/2}\cos\frac{\theta}{2}\ket{\downarrow}_\text{X}.
\end{equation}
Equation (\ref{eqn:EqualityCondition}) means that the operator
$V_\phi^\perp(\tau)\equiv\bras{\phi_\perp}{\text{X}}\times
e^{-iH_\text{tot}\tau}\ket{\phi}_\text{X}$ should have a zero
eigenvalue, and hence
\begin{equation}
\det V_\phi^\perp(\tau)=0.
\label{eqn:EqualityCondition2}
\end{equation}
For model (\ref{eqn:HamiltonianSingle}), the operator
$V_\phi^\perp(\tau)$ reads
\begin{widetext}
\begin{align}
V_\phi^\perp(\tau)
={}&\ketbras{\uparrow}{\uparrow}{\text{A}}
e^{-i(\Omega_\text{X}+\Omega_\text{A})\tau}\left[
1-e^{i(\Omega_\text{X}+\Omega_\text{A})\tau/2}\left(
\cos\delta\tau
+i\frac{\Omega_\text{X}-\Omega_\text{A}}{2\delta}\sin\delta\tau
\right)
\right]\sin\frac{\theta}{2}\cos\frac{\theta}{2}
\nonumber\displaybreak[0]\\
&{}-\ketbras{\downarrow}{\downarrow}{\text{A}}\left[
1-e^{-i(\Omega_\text{X}+\Omega_\text{A})\tau/2}\left(
\cos\delta\tau
-i\frac{\Omega_\text{X}-\Omega_\text{A}}{2\delta}\sin\delta\tau
\right)
\right]\sin\frac{\theta}{2}\cos\frac{\theta}{2}
\nonumber\displaybreak[0]\\
&{}+i\ketbras{\uparrow}{\downarrow}{\text{A}}
\frac{g}{\delta}e^{-i\varphi}
e^{-i(\Omega_\text{X}+\Omega_\text{A})\tau/2}
\sin\delta\tau\cos^2\!\frac{\theta}{2}
-i\ketbras{\downarrow}{\uparrow}{\text{A}}
\frac{g}{\delta}e^{i\varphi}
e^{-i(\Omega_\text{X}+\Omega_\text{A})\tau/2}
\sin\delta\tau\sin^2\!\frac{\theta}{2},
\end{align}
and
\begin{equation}
\det V_\phi^\perp(\tau)
=-\frac{1}{4}e^{-i(\Omega_\text{X}+\Omega_\text{A})\tau}\biggl[
\left|
1-e^{i(\Omega_\text{X}+\Omega_\text{A})\tau/2}\left(
\cos\delta\tau
+i\frac{\Omega_\text{X}-\Omega_\text{A}}{2\delta}\sin\delta\tau
\right)
\right|^2+\left(\frac{g}{\delta}\right)^2\sin^2\!\delta\tau
\biggr]\sin^2\!\theta.
\end{equation}
\end{widetext}
Condition (\ref{eqn:EqualityCondition}), namely
(\ref{eqn:EqualityCondition2}), is hence reduced to
\begin{subequations}
\begin{equation}
\sin\theta=0
\label{eqn:FirstPossibility}
\end{equation}
or
\begin{equation}
\cos\delta\tau=\pm1\quad\text{and}\quad
e^{i(\Omega_\text{X}+\Omega_\text{A})\tau/2}=\pm1.
\label{eqn:SecondPossibility}
\end{equation}
\end{subequations}

In the first case (\ref{eqn:FirstPossibility}), both conditions
(\ref{eqn:Condition}) and (\ref{eqn:OptimizationI}) are satisfied
unless $\delta\tau=n\pi$ ($n=1,2,\ldots$) as is explained around
(\ref{eqn:OptimizationSingle})--(\ref{eqn:OptimizationSingleII}).
In the second case (\ref{eqn:SecondPossibility}), on the other
hand, the projected time-evolution operator reads $V_\phi(\tau)
=\openone_\text{A}$ and the eigenvalue $\lambda_0=1$ is
degenerated, i.e., condition (\ref{eqn:Condition}) is not
fulfilled.
Therefore, the necessary and sufficient condition for the optimal
purification in model (\ref{eqn:HamiltonianSingle}) is given by
the first choice (\ref{eqn:FirstPossibility}) [i.e.,
(\ref{eqn:OptimizationSingle})] with
(\ref{eqn:OptimizationSingleII}).

\section{Condition for the Two-Qubit Initialization}
\label{app:EigenvaluesMultiple}
We here outline the proof of the necessary and sufficient
condition for the optimal two-qubit initialization,
Eq.~(\ref{eqn:InitializationCondition}), in
Sec.~\ref{sec:Initialization}.
What we have to show is how to make the eigenvalues $\lambda_\pm$
and $\lambda_{\uparrow\uparrow}$ in
(\ref{eqn:Initialize2Eigenvalues}) all less than unity in
magnitude.

The eigenvalues $\lambda_\pm$ are the solutions to an eigenvalue
equation
\begin{align}
(\lambda e^{i\Omega\tau})^2-2\left(
\cos^2\!\frac{\bar{g}\tau}{\sqrt{2}}
-\sin^2\!\chi\sin^2\!\frac{\bar{g}\tau}{\sqrt{2}}
\right)(\lambda e^{i\Omega\tau})&\nonumber\displaybreak[0]\\
{}+1-2\cos^2\!\chi\sin^2\!\frac{\bar{g}\tau}{\sqrt{2}}&=0.
\label{eqn:EigenvalueEqMultiple}
\end{align}
We clarify when this equation has a solution whose magnitude is
unity.
Seeking such a solution, we insert $\lambda=e^{-i\Omega\tau}
e^{i\Theta}$ into (\ref{eqn:EigenvalueEqMultiple}) to obtain the
conditions
\begin{subequations}
\begin{align}
\sin\Theta\left(
\cos\Theta
-\cos^2\!\frac{\bar{g}\tau}{\sqrt{2}}
+\sin^2\!\chi\sin^2\!\frac{\bar{g}\tau}{\sqrt{2}}
\right)&=0,\displaybreak[0]\\
\cos\Theta\left(
\cos\Theta
-\cos^2\!\frac{\bar{g}\tau}{\sqrt{2}}
+\sin^2\!\chi\sin^2\!\frac{\bar{g}\tau}{\sqrt{2}}
\right)&\nonumber\displaybreak[0]\\
{}-\cos^2\!\chi\sin^2\!\frac{\bar{g}\tau}{\sqrt{2}}&=0,
\end{align}
\end{subequations}
which are reduced to
\begin{subequations}
\begin{align}
\sin\frac{\bar{g}\tau}{\sqrt{2}}=0\quad&\text{and}\quad
\cos\Theta=1\displaybreak[0]\\
\intertext{or}
\cos\frac{\bar{g}\tau}{\sqrt{2}}=0\quad&\text{and}\quad
\cos\Theta=-1.
\end{align}
\end{subequations}
(Note that $\cos\chi\neq0$ and $\sin\chi\neq0$, since it is
assumed that $g_\text{XA},g_\text{AB}\neq0$.)
It is easy to see from (\ref{eqn:Initialize2EigenvaluesUpUp})
that we have $|\lambda_{\uparrow\uparrow}|=1$ when
$\sin(\bar{g}\tau/\sqrt{2})=0$, and in summary, the magnitude of
one of the eigenvalues $\lambda_\pm$ and
$\lambda_{\uparrow\uparrow}$ becomes unity only when
\begin{equation}
\sqrt{2}\bar{g}\tau=n\pi\quad(n=1,2,\ldots).
\end{equation}
The condition for the initialization in
Sec.~\ref{sec:Initialization} is thus proved to be
(\ref{eqn:InitializationCondition}).
See also Fig.~\ref{fig:EigenvaluesMultiple}.

\section{Condition for the Entanglement Purification}
\label{app:EigenvaluesEntanglement}
The necessary and sufficient condition
(\ref{eqn:ConditionEntanglement}) for the entanglement
purification in Sec.~\ref{sec:EntanglementPurification} is proved
in a similar manner to that in Appendix
\ref{app:EigenvaluesMultiple}.

The eigenvalues $\lambda_\pm$ and $\lambda_{\Phi^-}$ under the
condition (\ref{eqn:ConditionEntanglementI}) are the solutions to
an eigenvalue equation
\begin{align}
\left(\kappa-\sin\frac{g\tau}{\sqrt{2}}\right)\left(
\kappa-\cos^2\!\frac{\theta}{2}\sin\frac{g\tau}{\sqrt{2}}
\right)\left(
\kappa-\sin^2\!\frac{\theta}{2}\sin\frac{g\tau}{\sqrt{2}}
\right)\nonumber\displaybreak[0]\\
{}+2\sin^2\!\frac{\theta}{2}\cos^2\!\frac{\theta}{2}
\cos^2\!\frac{g\tau}{\sqrt{2}}\left(
\kappa-\frac{1}{2}\sin\frac{g\tau}{\sqrt{2}}
\right)=0
\end{align}
with $\kappa=(1-\lambda)/[2\sin(g\tau/\sqrt{2})]$.
Seeking a solution $\lambda$ with unit magnitude, we insert
$\lambda=e^{i\Theta}$ into this equation to obtain
\begin{subequations}
\begin{align}
&\sin\Theta\,\biggl[
2\cos^2\!\Theta
-\left(3-4\sin^2\!\frac{g\tau}{\sqrt{2}}\right)\cos\Theta
+1\nonumber\displaybreak[0]\\
&\phantom{\sin\Theta\,\biggl[}
{}-\left(2-\frac{1}{2}\sin^2\!\theta\right)
\sin^2\!\frac{g\tau}{\sqrt{2}}\left(
2-\sin^2\!\frac{g\tau}{\sqrt{2}}
\right)
\biggr]=0,\displaybreak[0]\\
&\cos\Theta\,\biggl[
2\cos^2\!\Theta
-\left(3-4\sin^2\!\frac{g\tau}{\sqrt{2}}\right)\cos\Theta
\nonumber\displaybreak[0]\\
&\phantom{\sin\Theta\,\biggl[}
{}-\left(2-\frac{1}{2}\sin^2\!\theta\right)
\sin^2\!\frac{g\tau}{\sqrt{2}}\left(
2-\sin^2\!\frac{g\tau}{\sqrt{2}}
\right)
\biggr]\nonumber\displaybreak[0]\\
&\quad\phantom{={}}
{}+1-2\sin^4\!\frac{g\tau}{\sqrt{2}}
-\frac{1}{2}\sin^2\!\theta\sin^2\!\frac{g\tau}{\sqrt{2}}\left(
2-3\sin^2\!\frac{g\tau}{\sqrt{2}}
\right)\nonumber\displaybreak[0]\\
&\quad=0,
\end{align}
\end{subequations}
which are reduced to
\begin{subequations}
\begin{align}
\sin\frac{g\tau}{\sqrt{2}}=0\quad&\text{and}\quad
\cos\Theta=1\displaybreak[0]
\label{eqn:EntanglementNoGoI}
\displaybreak[0]\\
\intertext{or}
\cos\frac{g\tau}{\sqrt{2}}=0\quad&\text{and}\quad
\cos\Theta=-1.
\label{eqn:EntanglementNoGoII}
\end{align}
\end{subequations}
Extraction of entanglement is not possible when
(\ref{eqn:EntanglementNoGoI}) or (\ref{eqn:EntanglementNoGoII})
is satisfied, and therefore, the condition for the entanglement
purification in Sec.~\ref{sec:EntanglementPurification} is given
by (\ref{eqn:ConditionEntanglement}).



\begin{thebibliography}{10}
\bibitem{note:QZE}
It should be noted, however, that the time interval $\tau$ in
this scheme is not necessarily small as in the ordinary Zeno
measurements, and the purification (\ref{eqn:Purification}) is
not due to the quantum Zeno effect.  If the ordinary Zeno limit
$N\to\infty$ and $\tau\to0$ ($N\tau=T$ fixed) \cite{ref:QZE}
is taken in the present scheme, a quantum Zeno effect appears
yielding the so-called ``quantum Zeno dynamics'' \cite{ref:QZD},
which is unitary and provides us with a quite different effect
from the one discussed in this article.
\bibitem{note:Purification}
One should keep in mind, however, that some other schemes
\cite{ref:QuantInfoCompZeilinger,ref:PurificationBennett} are
discussed in more severe situations.
\end{thebibliography}
\end{document}